\documentclass[a4paper]{jpconf} 
\usepackage{graphicx}
\begin{document}
\title{Four Aspects of Superoscillations}

\author{Achim Kempf}

\address{Departments of Applied Mathematics and Physics\\ Institute for Quantum Computing\\ University of Waterloo \\ Waterloo, N2L 3G1, Ontario, Canada}

\ead{akempf@uwaterloo.ca}

\begin{abstract}
A function $f$ is said to possess  superoscillations if, in a finite region, $f$ oscillates faster than the shortest wavelength that occurs in the Fourier transform of $f$. I will discuss four aspects of superoscillations: 1. Superoscillations can be generated efficiently and stably through multiplication. 2. There is a win-win situation in the sense that even in circumstances where superoscillations cannot be used for superresolution, they can be useful for what may be called superabsorption, an effective up-conversion of low frequencies 3. The study of superoscillations may be useful for generalizing the Shannon Hartley noisy channel capacity theorem. 4. The phenomenon of superoscillations naturally generalizes beyond bandlimited functions. \end{abstract}

\noindent \bf Introduction. \rm The phenomenon of superoscillations was first discussed in pioneering works that include, in particular, \cite{earliest-1}-\cite{beethoven}. In the meantime, various methods for generating and analyzing superoscillatory wave forms in theory and practice have been developed, see, e.g. \cite{theory-1}-\cite{theory-l}, and various possible applications are being explored, see, e.g., \cite{apps-1}-\cite{apps-l}, in particular, for the purpose of superresolution. Here, I discuss four recent developments and prospects in the field of superoscillations.

\section{Superoscillations can be generated efficiently and stably through multiplication.}
Let us consider the problem of constructing a function, $f$, that obeys a given bandlimit, $\Omega$, and that is superoscillating in the sense that it has $N$ zero crossings, $f(t_n)=0$, at a set of points, $\{t_n\}_{n=1}^N$, that are  arbitrarily more densely spaced than the Nyquist spacing.   

As we showed in \cite{leilee}, it is possible to generate such superoscillations efficiently and stably by multiplication. The new method is based on the simple fact that for the product of bandlimited functions, the bandwidth is additive. In particular, assume that each of a set of $N$ functions $\{g_n\}_{n=1}^N$ is $\Omega/N$-bandlimited. Then, their product
\begin{equation}
    f:=\prod_{n=1}^N g_n
\end{equation}
is a function with bandlimit $\Omega$. Further, the function $f$ possesses all the zeros that the functions $g_n$ possess. 

Now let us choose the $\Omega/N$-bandlimited functions $g_n$, in such a way that $g_n(t_n)=0$ for $n=1,...,N$. This is simple because we can translate any $\Omega/N$-bandlimited function along the $t$-axis so that one of its zeros is at $t_n$ without affecting the bandlimit. For example, we can choose the functions $g_n$ to be suitable translates of another. As a result, we obtain that the function $f$ is $\Omega$-bandlimited and superoscillates as desired because it possesses all the prescribed zero crossings. 

The new method has certain advantages, in particular, when compared to the additive method for generating superoscillations \cite{beethoven}. There, superoscillatory functions are constructed as a linear combination of $\Omega$-bandlimited functions. The coefficients of such linear combinations are generally extremely fine tuned and need to be calculated by inverting an ill-conditioned matrix. This implies an inherent need for extreme numerical accuracy. The multiplicative method avoids these problems. 

Second, in \cite{theory-1,ak-fe1,ak-fe2}, using the additive method, it was shown that the dynamic range of superoscillatory functions, i.e., the ratio of the largest amplitudes (which are related to the $L^2$ norm due to the finiteness of $N$) to the superoscillatory amplitudes, obey certain scaling laws: the dynamic range grows polynomially with the frequency of the superoscillations and it grows exponentially with the number of superoscillations. 
The new multiplicative method makes it clear how these scaling laws arise: we can assume that the prescribed zeros are much closer than the Nyquist spacing. This means that the functions $g_n$ can be assumed to be well approximated in the superoscillating region as a low order or even first order polynomial about their zero crossing point $t_n$. The amplitudes of the function $f$ in the superoscillating region are, therefore, exponentially suppressed with $N$ and polynomially (to the power of $N$) suppressed with the spacing of the $t_n$. For the detailed calculation, see \cite{leilee}.

The new multiplicative method for generating superoscillatory wave forms invites experimental implementation wherever signals can be multiplied, as is the case, for example, in electronics where transistors can be used to multiply signals. 

\section{Superoscillations for superresolution or superabsorption: a win-win situation.}
In order to detect landmines via radar from the air, a spatial resolution on the order of the size of landmines, i.e., of the order of $10^{-1}m$, would be required. However, electromagnetic waves of such wavelengths are very efficiently absorbed by even small amounts of humidity on top of or in the soil. Radar waves of significantly longer wavelengths would penetrate the ground but then the spatial resolution necessary to detect landmines would be expected to be lost. Let us discuss if the use of superoscillatory radar signals could be useful in this case. 

The initial idea to explore here, \cite{PPA}, is that a superoscillatory radar pulse of sufficiently small bandwidth should penetrate the ground while the radar pulse's superoscillatory stretch could be used to enhance the resolution to the level needed to discern the landmines. 

However, in this example of how superoscillations might be used to achieve superresolution a crucial assumption is made, an assumption which may not hold in practice. The assumption is that each water molecule interacts  with the entire radar pulse without being disturbed. This assumption is important because if the water molecule is being disturbed, namely by interacting with its environment while the radar pulse passes, then the water molecule might only see time-windowed portions of the passing radar signal. In this case, the water molecule can resonate with the superoscillations in the passing radar pulse. This is because while the temporarily high frequencies in a superoscillatory function are not present in the superoscillatory function's full Fourier transform, these high frequencies are present in windowed Fourier transforms of the superoscillatory signal. The water molecule can, therefore, absorb energy from the radar pulse and quickly dissipate this energy into its environment. 

In fact, as we showed in \cite{angus1,angus2}, a superoscillatory pulse will excite an oscillator (or a water molecule) in interesting ways that make this dissipative process possible. To see this, let us begin with the observation that if an $\Omega$-bandlimited function, $f$, is superoscillating in an interval $[t_1,t_2]$, then $f$ contains the exact  same total amount of superoscillations also outside of this interval. Concretely, if the Fourier transform of $f$ over just the interval $[t_1,t_2]$ contains a frequency $\omega>\Omega$ with amplitude $a_\omega$, then the Fourier transform of $f$ over the outside of the interval $[t_1,t_2]$ must also contain this frequency $\omega$, namely with the opposite amplitude $-a_\omega$. This is because by assumption, the signal $f$ is $\Omega$-bandlimited, i.e., the Fourier transform of $f$ over the entire real line must contain any frequency $\omega>\Omega$ with amplitude $0$.   

As a consequence, as was shown in \cite{angus1,angus2}, a superoscillating signal as an external driving force will start to resonantly excite an oscillator right away, until the starting time of the superoscillatory stretch, $t_1$, which is when the signal will start to resonantly de-excite the oscillator. The de-excitation of the oscillator is complete at time $(t_2-t_1)/2$ when the signal starts exciting the oscillator again, up until time $t_2$. From time $t_2$ until $t\rightarrow\infty$, the signal then de-excites the oscillator. Eventually, the signal will, therefore, leave the oscillator asymptotically unexcited, as it should be because, by assumption, the full signal did not possess the resonance frequency of the oscillator. In the above, we assumed, for simplicity that the superoscillatory's signal is symmetric about the time $(t_2-t_1)/2$.  

It is clear that if these two excitation/de-excitation processes are disturbed because the oscillator, or the water molecule, is interacting with its environment, then the oscillator or water molecule will generally dissipate some energy that it absorbed from the superoscillatory signal into its environment. Of course, the process by which water molecules absorb electromagnetic waves of wavelength of a few centimeters (mostly in their rotational degrees of freedom) and handing this energy to its neighbors is the principle of microwave ovens. The question is whether this process is efficient also with signals that do not possess the resonance frequencies but that only superoscillate at a resonance frequency. To this end, it is crucial to compare the scale of the typical time, $\Delta t_{collision}$, between interactions of water molecules with their environment and the duration, $\Delta t_{pulse}$, of the superoscillating radar pulse.  

We can crudely estimate that $\Delta t_{collision}\approx(10^{-9}m)/(10^3m/s)=10^{-12}s$, given that the typical distance between molecules is on the order of $10^{-9}m$ and that their typical velocity is on the order of the speed of sound in water, $10^3m/s$. A crude estimate of the duration of a superoscillating radar pulse 
is $\Delta t_{pulse}\approx (1m)/(10^8m/s)=10^{-8}s$, given the speed of light at the order of $10^8m/s$ and given that the superoscillating pulse needs to possess a length on the order of a meter in order to contain both, a few superoscillations of say tens of centimeters of length plus the large amplitudes before and after the superoscillating stretch. 

This rough estimate shows that $\Delta t_{collision} \ll  \Delta t_{pulse}$, i.e., we should expect that during that a radar pulse passes, a water molecule will repeatedly interact with its neighbors. As a consequence, the superoscillatory radar signals should not penetrate the ground undisturbedly, so that superresolution of land mines at significant depths should be difficult to achieve in this manner. However, while this leaves landmine detection difficult, this example also shows that in scenarios where superresolution does not work, there are other opportunities. 

Namely, in any situation where superresolution is hindered in this way by absorption due to too fast interactions in the medium, the superoscillatory signals do offer a method to probe for exactly such subtle absorption processes that are mediated by very fast processes in media. In this sense, superoscillatory signals can give us either a desired high resolution in space, or in a different way, a potentially useful high sensitivity to very fast absorption processes.  

For an example, let us consider the promising still relatively new field of optogenetics, see, e.g., \cite{opto-1}-\cite{opto-l}. Also in optogenetics, relatively short wavelengths tend to be needed while only longer wavelengths can penetrate the medium. Recall that in the field of optogenetics, living cells, such as neurons, are genetically modified to express proteins and ion channels that are light sensitive. This enables the time and space resolved recording and active control of certain processes in living organisms through light. Optogenetics, thereby, offers the prospect of a new avenue, for example, for the treatment of neurological disorders. 

One bottleneck of the optogenetic approach is the fact that optogenetic methods normally require the use of blue light. As blue light is readily absorbed by biological tissue, optogenetic applications deep inside biological tissue, such as brain tissue, tend to require invasive methods such as the insertion of optical fibres to supply blue light to regions deep in the tissue, see, e.g., \cite{opto-a}. A recent approach to overcoming this problem is to inject frequency upconverting nanoparticles into the deep tissue area and then irradiate the tissue from the outside with near-infrared light (NIR) that is of sufficiently long wavelength to easily penetrate the tissue, see \cite{opto-b}. 

This suggests the use of superoscillatory NIR radiation for the purpose of sending electromagnetic energy deep into tissue where then a superoscillating stretch might interact with the genetically modified cells. Leaving aside for now the question of how to generate superoscillating NIR signals in practice, let us again make a rough estimate of the orders of magnitude involved.  

In this case, in tissue as in soil, we again crudely estimate that thermal collisions of molecules happen roughly at time intervals of order $\Delta t_{collision}=10^{-12}s$. The length of a superoscillating NIR pulse could be of the order of say tens of NIR wavelengths (micrometers), i.e., of the order of $10^{-5}m$. This yields for the duration of a superoscillating NIR pulse the order of $10^{-5}m/(10^8m/s)=10^{-13}s$. This means that the relative sizes of the pulse duration and the typical collision time could be favorable and it should, therefore, be interesting to explore the possible use of superoscillatory NIR radiation in optogenetics. 

\section{Superoscillations and Shannon's noisy channel capacity theorem.}

One of the reasons why the existence of superoscillatory bandlimited functions is surprising is that their existence appears to contradict the Shannon Hartley theorem \cite{thomascover}. Let us recall the precise way in which the theorem establishes an upper limit to the rate at which information can be transmitted through a bandlimited noisy channel. Assume that the bandlimit of the channel is $\Omega$, the average signal power is $S$ and the average noise power is $N$. Let us assume, further, that the noise is additive Gaussian white noise. The Shannon Hartley theorem states that then the channel capacity, $C$, i.e., the maximum rate at which information can be reliably transmitted through the channel, in bits per unit time, is given by: 
\begin{equation}
    C= \Omega \log_2\left(1+\frac{S}{N}\right) \label{shannonhartley}
\end{equation}
The channel capacity, therefore, grows logarithmically with the signal to noise ratio $S/N$ while growing linearly with the bandwidth, $\Omega$. 

The existence of superoscillations among $\Omega$-bandlimited signals appears to challenge the Shannon Hartley theorem. This is because it would appear that by using superoscillations, one may be able to make a bandlimited signal oscillate arbitrarily fast in prescribed ways, independently of the bandwidth $\Omega$ in Eq.\ref{shannonhartley}. This suggests that it may be possible to pack information into such a signal at an arbitrarily high rate, which would be a contradiction to Eq.\ref{shannonhartley}. Let us now study the way in which superoscillatory signals avoid contradicting the Shannon Hartley theorem. 

To see this, let us begin by recalling that, as was proven in \cite{beethoven}, among $\Omega$-bandlimited signals, one can always find a signal, $f(t)$, which passes through an arbitrary number, $N$, of arbitrarily prescribed points $t_n, n=1,...,N$. Concretely, assume given a bandlimit $\Omega$. Further, choose $N$ arbitrary real-valued sample times $t_n, n=1,...,N$ and real-valued amplitudes $a_n, n=1,...N$. It is always possible to construct an $\Omega$-bandlimited function that obeys:
\begin{equation}
    f(t_n) = a_n, ~~~~\forall~n=1,...N \label{shannon}
\end{equation}
For example, it is possible to find a $1Hz$-bandlimited function that passes through, say, $10^{20}$ amplitudes of a one-hour Beethoven symphony recorded with a 20KHz bandlimited system. How can this fact be consistent with the Shannon Hartley theorem? It is consistent because such a strongly superoscillatory signal must possess extremely large amplitudes outside the interval in which the superoscillations occur. In fact, as was shown in \cite{theory-1}-\cite{ak-fe2}, the signal's $L^2$ norm must grow exponentially with the number of superoscillations. Further, it was shown that this exponentially large $L^2$ norm arises not from a slow decay towards $\vert t\vert\rightarrow \infty$ but from exponentially large  amplitudes in the neighborhood of the superoscillating stretch. 

It is important that these large amplitudes outside the superoscillatory interval cannot be removed from the signal without removing its property of $\Omega$-bandlimitation. Only the entire signal, including its large amplitudes outside the superoscillatory stretch is $\Omega$-bandlimited and will, therefore, be able to pass through the $\Omega$-bandlimited channel. This means that the price to pay to be able to send a 1Hz-bandlimited superoscillating Beethoven signal through a 1Hz bandlimited channel is to have to send all of this signal, including its exponentially large amplitudes outside the superoscillating interval. This, however, requires that the dynamic range of the channel must be such that it can transmit both the amplitudes of the superoscillating Beethoven stretch as well as the exponentially larger amplitudes outside that stretch. For the Beethoven stretch to be at or above the noise level of the channel, the overall average signal amplitude, which is of course dominated by the large amplitudes outside the superoscillating stretch, must be exponentially large. 

We obtain, therefore, that by using superoscillations it is possible to transmit information at a rate that is arbitrarily larger than one may expect given the bandwidth of the signal. However, the price that is to be paid, and which makes superoscillations consistent with the Shannon Hartley theorem, is that the signal-to-noise ratio $S/N$ needs to grow exponentially with the number of superoscillations - consistent with Eq.\ref{shannon}. 

Interestingly, the consistency of the existence of arbitrarily highly superoscillatory signals with the Shannon Hartley theorem points towards an opportunity to widely generalize the Shannon Hartley theorem. To see this, let us recall that the Shannon Hartley theorem requires the assumption of a particular model for the noise in the channel, namely Gaussian white noise. Any other noise model requires the derivation of a corresponding new version of the Shannon Hartley theorem. 

In contrast, the properties of superoscillations do not depend on any particular noise model. The key property of superoscillatory signals, namely that what is in effect their dynamic range grows exponentially with the number of superoscillations (and polynomially in the frequency of the superoscillations), see \cite{theory-1}-\cite{ak-fe2} and \cite{leilee}, does not refer to a noise model. Instead, what plays the role of $S/N$ from the perspective of  superoscillatory signals is the dynamic range in the sense of the ratio of the large amplitudes outside the superoscillating stretch to the small amplitudes of the superoscillations. 

This suggests the conjecture that a generalized Shannon Hartley theorem can be proven which uses the generic scaling properties of superoscillations to relate the channel capacity linearly to the bandwidth and logarithmically not to $S/N$, which is a noise-model dependent quantity, but instead to what is in effect the dynamic range of superoscillations. The challenge is to find a natural and robust definition of this dynamic range along with possibly a suitable generalization of the notion of channel capacity. The benefit of such a theorem would be a generalized channel capacity theorem that applies independently of particular noise models. Such a theorem could give a concrete information theoretic meaning to the observation that superoscillatory stretches of arbitrary finite frequencies and duration are possible and that the price to pay is a dynamic range that grows exponentially with the number of superoscillations and polynomially with their frequency. 

\section{Superoscillations beyond bandlimited functions.}

The phenomenon of superoscillations generalizes beyond bandlimited functions. To see this, let us consider $N$ generic functions $b_n, n=1,...,N$. Those functions can be chosen to be bandlimited but they do not need to be bandlimited. The functions $b_n$ linearly span a function space, ${\cal F}:= \mbox{span}\left(\{b_n\}_{n=1}^N\right)$:
\begin{equation}
f(t) = \sum_{n=1}^N f_n b_n(t)~~~~~    \forall f\in {\cal F}  
\end{equation}
Does ${\cal F}$ contain functions that oscillate arbitrarily quickly? The answer is yes, in the sense that we can generally prescribe arbitrary amplitudes $a_i$ at up to $N$ points $t_i$ and there will exist a function $f\in {\cal F}$ that passes through those $N$ points. For example, let us ask if there exists a function $f\in {\cal F}$ that obeys $
    f(t_m) = (-1)^m~~~\mbox{for}~~~~ m=1,2,...,N
$
even if we choose the points $t_m$ arbitrarily close together. To this end, we need to solve
\begin{equation}
    f(t_m) = a_m=(-1)^m = \sum_{n=1}^N f_n b_n(t_m) \label{sum}
\end{equation}
for the coefficients $f_n$. Rewriting Eq.\ref{sum} as the vector equation $\vec{a}=B\vec{f}$ with the matrix 
\begin{equation} B_{mn}:=b_n(t_m),
\end{equation} 
we obtain that there exists a solution,
\begin{equation}
    \vec{f}=B^{-1}\vec{a}
\end{equation}
if and only if $B$ is invertible, i.e., if the determinant of $B$ is nonzero, det$(B)\neq 0$. If the functions $b_n$ are chosen randomly, the probability that this determinant vanishes is zero. 

We remark that in the special case where the functions $b_n$ are bandlimited functions, we recover conventional superoscillations. It was shown in \cite{theory-1}-\cite{ak-fe2} that if the aim is to prescribe amplitudes $a_n$ at an arbitrary number, $N$, of points, $t_m$, then for a given bandlimit $\Omega$, the optimal choice of the $N$ basis functions $b_n$ is to choose the $N$ $\Omega$-bandlimited sinc-functions that are centered at the points $t_n$. This choice is optimal in the sense that the resulting superoscillating functions possess the smallest possible $L^2$ norm among all $\Omega$-bandlimited functions that pass through the prescribed points. 

While the above calculation provides a proof for the existence of superoscillations in generic function spaces, the method that it provides for constructing such superoscillatory functions suffers from a technical difficulty when applied in practice. This is because the method requires the calculation of the inverse of the matrix $B$. When the points $t_m$ are chosen close together, $B$ can become very ill-conditioned, see \cite{theory-1}-\cite{ak-fe2}, which makes this matrix inversion numerically unstable. In practice, it can be preferable, therefore, to construct superoscillations multiplicatively, as explained in Sec.1 above, in which case no matrix inversion is required. 

Further, for the above proof of the existence of superoscillations, the points $t_m$ need not necessarily be real numbers but can also be points on an arbitrary differentiable manifold, ${\cal M}$, since this simple calculation applies to spaces of functions $f:~I\!\!R\rightarrow I\!\!R$ as well as to spaces of real (or complex-valued) functions on arbitrary differentiable manifolds, $f:~{\cal M} \rightarrow I\!\!R$. 

This suggests the question as to the significance of superoscillations for physical fields in curved spacetime. The notion of bandwidth  can be generalized straightforwardly to Riemannian manifolds. The space of covariantly $\Omega$-bandlimited functions then consists of the space of eigenfunctions of the Laplacian on 0-forms whose eigenvalues are below $\Omega^2$. A generalization to Lorentzian manifolds is obtained by instead cutting off the spectrum of the d'Alembert operator, see \cite{ak-1}-\cite{ak-2}. The significance of superoscillations in quantum field theory has been studied, in particular, in the case of the free Klein Gordon field in 1+1 dimensional spacetime with a bandlimitation in the form of a cutoff on the momentum \cite{jason}. A key finding has been that as the $N$ points $t_n$ are moved closer together than the Nyquist spacing, the vacuum entanglement entropy of the effectively superoscillating region remains of the order of the entanglement entropy of the region covered by $N$ points at Nyquist spacing. This confirms for the case of quantum field theory the intuitive expectation that, in spite of the existence of superoscillations, it is the Nyquist spacing which represents the density of degrees of freedom. 

\ack
This work has been supported by the Discovery Grant Program of the National Science and Engineering Research Council of Canada (NSERC).

\end{document}